\documentclass[12pt]{article}
\usepackage{amsfonts}
\usepackage{graphicx}

\newcommand{\eq}{\begin{equation}}
\newcommand{\eqx}{\end{equation}}
\newcommand{\eqn}{\begin{eqnarray}}
\newcommand{\eqnx}{\end{eqnarray}}
\newcommand{\f}[2]{\frac{#1}{#2}}
\newcommand{\qqqq}{\quad\quad\quad\quad}

\newcommand{\nn}{{\cal N}}

\newcommand{\dl}{\delta}

\newcommand{\al}{\alpha}
\newcommand{\bt}{\beta}
\newcommand{\eps}{\varepsilon}
\newcommand{\cor}[1]{\left\langle{#1}\right\rangle}

\newcommand{\tr}{\mbox{\rm tr}\,} 

\newcommand{\zfg}{z_{FG}}

\makeatletter
\@addtoreset{equation}{section}
\def\theequation{\thesection .\arabic{equation}}
\makeatother

\title{From static to evolving geometries -- R-charged hydrodynamics
  from supergravity} 

\author{ Dongsu Bak$\,^{a}$ and Romuald A. Janik$\,^b$\thanks{
e-mails: {\tt dsbak@mach.uos.ac.kr}, {\tt ufrjanik@if.uj.edu.pl}}\\ \\
$^a$ \small \it Physics Department, University of Seoul,\\ 
\small \it Seoul 130-743, Korea \\
$^b$ \small \it Institute of Physics, Jagellonian University,\\
\small \it Reymonta 4, 30-059 Krakow, Poland.}

\date{}

\begin{document}

\maketitle

\begin{abstract}
We show that one can obtain asymptotic evolving boost-invariant
geometries in a simple manner from the corresponding static
solutions. We exhibit the procedure in the case of a
supergravity dual of R-charged
 hydrodynamics by turning on a supergravity
gauge field and analyze the relevant thermodynamics. Finally we consider
turning on the dilaton and show that electric and magnetic modes in
the plasma equilibrate before reaching asymptotic proper times.   
\end{abstract}

\section{Introduction}

Quark-gluon plasma, the deconfined state of matter produced at RHIC is
currently the focus of numerous experimental and theoretical
studies. Since it appears that the gauge theory coupling in the plasma
is large \cite{largegym}, it is interesting to develop nonperturbative
methods for studying its properties.

A very effective framework for studying nonperturbative phenomena is
the AdS/CFT correspondence \cite{adscft}. Even in its simplest form,
for $\nn=4$ Super-Yang-Mills theory, at finite temperature one may
expect many similarities with properties of strongly coupled
deconfined QGP.

A lot of work has been done in the study of properties of static
plasma at fixed temperature e.g. transport coefficients have been
computed \cite{son,other}, the drag force acting on a moving quark was
investigated \cite{dragforce}. Much less is known regarding questions
related to time-dependent phenomena like the origin of thermalization,
hydrodynamic evolution etc. Works investigating these issues include
\cite{nastase,zahed,JP1,JP2,SJSin,RVisc}.  

In order to gain more understanding of the above mentioned dynamical
processes in strongly interacting gauge theory using AdS/CFT it is
worthwhile to investigate the structure of dual time-dependent
geometries. In general holographic renormalization \cite{skenderis} gives a
prescription for constructing a dual geometry to {\em any} given gauge
theory energy-momentum spacetime profile. In \cite{JP1} it was
advocated that the requirement of {\em nonsingularity} of the dual
geometry selects the physical energy-momentum evolution in the gauge
theory. At leading order, in a boost invariant setting, this
requirement led to late-time perfect fluid hydrodynamical evolution
\cite{JP1}, while carrying out the analysis also for subleading asymptotic
times\footnote{And using some results of \cite{SJSin}.
}~\cite{RVisc}
determined the effects of viscosity on the evolution with the exact
(shear) viscosity coefficient.

The dual geometry for late asymptotic times constructed in \cite{JP1}
bears a remarkable similarity to the static black hole but with the
position of the horizon moving with a specific scaling with
proper-time. In this note
we would like to perform an analogous construction for the more
complicated case with R-charged matter, show that a similar
phenomenon also occurs in this case and exhibit the origin of such a
behavior. 

Another direction of generalization of the evolving geometry is to
turn on the dilaton field. This has the physical interpretation on the
gauge theory side of allowing for differing expectation values of
squares of
electric and magnetic fields in the evolving plasma. We find that, for
asymptotic proper-times, in order to have a nonsingular geometry one
has to have equilibration between electric and magnetic modes.

The plan of this paper is as follows. First in section 2 we will briefly
review boost-invariant hydrodynamics with a conserved charge. In
section 3, we will construct the dual charged evolving geometry for
asymptotic proper-times and explain the origin of its marked similarity
to the corresponding static geometry. 
We will then examine, in section 4, the thermodynamics of the
resulting evolving system. Finally, in section 5, we will consider
turning on the dilaton. We close the paper with concluding remarks.

\section{Hydrodynamics with a conserved charge}

Perfect fluid hydrodynamics with a conserved charge is described by
the energy-momentum tensor and the current
\eq
T^{\mu\nu}=(\eps+p)u^\mu u^\nu +p \eta^{\mu\nu}\,, \qqqq 
J^\mu= \rho u^\mu\,,
\eqx
which are conserved i.e.
\eq
\partial_\mu T^{\mu\nu}=0\,, \qqqq \partial_\mu J^\mu=0\,.
\eqx
Moreover for the conformal theory that we are considering
$T^\mu\!_\mu=0$ which gives $\eps=3p$.

Let us now impose the requirement of boost-invariance in the
longitudinal plane and no dependence on transverse coordinates. In the
natural proper-time/spacetime rapidity coordinates the Minkowski metric
has the form
\eq
ds^2=-d\tau^2+\tau^2 dy^2 +dx_\perp^2\,.
\label{bdmet}
\eqx
Boost-invariance then forces the fluid velocity to be
$u^\mu=(1,0,0,0)$, and energy momentum conservation leads to Bjorken
evolution,
\eq
\eps(\tau)=\f{e_0}{\tau^\f{4}{3}}\,,
\eqx
while current conservation leads to
\eq
\label{cons}
\nabla_\mu J^\mu 
=
\f{dJ^\tau}{d\tau} +\f{1}{\tau} J^\tau =0
\quad\quad \longrightarrow \quad\quad J^\mu=(\f{\rho_0}{\tau},0 ,0,0)\,.
\eqx
We will show below that this scaling leads to a nonsingular dual
geometry.

\section{Supergravity analysis}

The 5D Einstein Maxwell action for the the minimal cases
is given by
\begin{eqnarray}
I= {1\over 16\pi G_5}
\int \left(\sqrt{-g} \Bigl(R+ 12 - {1\over 4} F_{\alpha\beta}
F^{\alpha\beta}\Bigr) - \alpha \epsilon^{\alpha\beta \mu\nu\lambda}
F_{\alpha\beta} F_{\mu\nu} A_\lambda
\right)\,.
\end{eqnarray}
where $G_5 = \pi/(2N_c^2)$ and 
$\alpha=1/(48\sqrt{3})$ in our convention.
Any solution of the above action can be consistently embedded
into the 10D type IIB supergravity \cite{Chamblin}.
The gravity equations of  motion become
\begin{eqnarray}
\label{einstein}
&& R_{\alpha\beta}= -4 g_{\alpha\beta} -{1 \over 12} {F_{\mu\nu} 
F^{\mu\nu}}\,
g_{\alpha\beta}  + {1\over 2} F_{\alpha}\,^\mu F_{\beta\mu} 
\,,\\
&& \nabla_\alpha F^{\alpha\mu} - {3\alpha \over \sqrt{-g}} 
\epsilon^{\mu\alpha\beta\gamma\delta} F_{\alpha\beta} 
F_{\gamma\delta} 
=0\,.
\end{eqnarray}

The static black hole solution dual to the finite temperature $\nn=4$ SYM
plasma with R-charge chemical potential is given by
\begin{eqnarray}
\label{metric}
ds^2 &=& {1\over z^2} (-h(z) dt^2 + d\vec{x}^2 + dz^2/h(z) )
\ , \\
F_{zt}&=& q z\ ,
\end{eqnarray}
where $h(x)= 1- a x^4 + q^2 x^6/12$ and all other components of the gauge 
field are vanishing.

In order to analyze the gauge theory energy-momentum tensor in the
most convenient manner one usually passes to the Fefferman-Graham
coordinates \cite{skenderis}. However it turns out that contrary to
the uncharged case $q=0$ considered in \cite{JP1}, the Fefferman-Graham   
form of (\ref{metric}) cannot be expressed in terms of elementary
functions. Therefore it is more convenient to adopt coordinates
similar to (\ref{metric}) also for the evolving case.

Namely the proper-time dependent metric ansatz will be
\begin{eqnarray}
ds^2 &=& {1\over z^2} (- e^{A(z,\tau)} d\tau^2 + 
\tau^2 e^{B(z,\tau)}  dy^2 + dx_\perp^2  
+ e^{D(z,\tau)}    dz^2 )
\ , \\
F_{z\tau}&=& K(z,\tau)\ ,
\end{eqnarray}
which is a general ansatz compatible with boost invariance. Note that with
respect to the Fefferman-Graham form of coordinates, the $z$ and $\tau$
coordinates have to be redefined. With this ansatz, the gauge field
equation part can be solved with an integration constant $q$ by
\begin{equation}
F_{z\tau} = q {z\over \tau} e^{\f{A+D-B}{2}}\,.
\label{maxsol}
\end{equation} 
Then the scalar $F^2$ becomes
\begin{equation}
F^2= - 2 q^2 {z^6\over \tau^2} e^{- B}\,.
\label{fsquare}
\end{equation} 
To study nontrivial late time scaling behaviors, we take the scaling 
variable as $v= z/\tau^{s/4}$ and work in the large $\tau$ limit
while keeping $v$ fixed.  
Then in the scaling limit,  one has
\begin{eqnarray}
\!\!\!\!A \!\!&=&\!\! a(v) + O(1/\tau^\natural)\,, \ \ 
B \!\!=\!\! b(v) + O(1/\tau^\natural)\,, \ \ 
D \!\!=\!\! d(v) + O(1/\tau^\natural)\,,\\
\!\!\!\!F^2 \!\!&=&\!\!  f^2 (v) +O(1/\tau^\natural)\ ,
\end{eqnarray}
where $\natural$ denotes some positive power so that
the terms can be ignored in the large proper-time limit. One can, of course,
consider the case where $f^2=0$ but then the analysis 
will be reduced to the problem of uncharged fluid dynamics.

From (\ref{fsquare}), one concludes then that, in the scaling variable 
$v= {z\over \tau^{s/4}}$,  the power $s$ should be fixed as
$4/3$ once we have a nonvanishing charge density in the 
scaling limit. 

Let us rewrite the Einstein equation (\ref{einstein}) in the following
form
\eq
R^\al\!_\bt=-\dl^\al\!_\bt \left(4+\f{1}{12} F^2\right) +\f{1}{2}
F^{\al\mu} F_{\bt\mu}\,.
\eqx
The scaling large $\tau$ limit of the diagonal equations
($\tau\tau$, $zz$, $yy$ and $xx$ components) is
\eqn
\label{gtt}
R^\tau\!_\tau = -4+\f{1}{6} F^2 &\qqqq&
R^z\!_z = -4+\f{1}{6} F^2 \\
R^y\!_y = -4-\f{1}{12} F^2  &\qqqq&
\label{gxx}
R^x\!_x = -4-\f{1}{12} F^2\,,
\eqnx
where the explicit formulas for the $R^\al\!_\bt$ in the scaling limit
are given in the appendix. 
Apart from these equations we also have the leading part of the
off-diagonal equation $R^\tau\!_z=0$:
\eq
\label{rtz}
v(2b''(v)+b'(v)^2)+2(3a'(v)+3d'(v)-2b'(v))-v(a'(v)+d'(v))b'(v)=0\,.
\eqx

Remarkably enough a direct analog of the static solution:
\eqn
a(v) &=& \log\left(1-av^4+\f{q^2}{12} v^6 \right) \\
b(v) &=& 0 \\
d(v) &=& -\log\left(1-av^4+\f{q^2}{12} v^6 \right)
\eqnx
solves all the above equations! If we recall now the definition of the
scaling variable $v=z/\tau^{1/3}$, we see that asymptotically for
large $\tau$ the geometry looks like the static charged black hole
with
\eq
h=1-a_{eff}(\tau) z^4 +\f{q^2_{eff}(\tau)}{12} z^6
\eqx
but with the parameters $a_{eff}(\tau)$ and $q_{eff}(\tau)$ being
$\tau$-dependent. In particular the scaling is exactly such that the
effective charge behaves like
\eq
q_{eff}(\tau) \sim \f{q}{\tau}
\eqx
as follows from (\ref{cons}). We shall return to the more precise 
determination of
charge density later on.  

Before we examine in detail the physical properties of the evolving
solution let us try to understand why such a simple generalization of
the static solution gives an evolving boost invariant solution. 

It turns out that the diagonal components of the Ricci tensor
$R^\al\!_\bt$ computed for the {\em static} metric
\eq
ds^2=\f{1}{z^2} \left( -e^{a(z)}dt^2+e^{b(z)} dy^2+e^{c(z)} dx_\perp^2
+e^{d(z)} dz^2 \right)
\eqx
coincide with the scaling limit of diagonal components of the Ricci
tensor $R^\al\!_\bt$ of the boost-invariant {\em evolving} metric
\eq
ds^2=\f{1}{z^2} \left( -e^{a(v)}d\tau^2+e^{b(v)} 
\tau^2 dy^2+e^{c(v)} dx_\perp^2
+e^{d(v)} dz^2 \right)
\eqx
with the substitution of $z$ for $v=z/\tau^{1/3}$. A similar scaling
property holds here for the gauge field, hence equations
(\ref{gtt})-(\ref{gxx}) for the evolving case coincide with the
corresponding equations for the static metric with $v$ interchanged
with $z$.

Once equations (\ref{gtt})-(\ref{gxx}) are solved,
the off-diagonal equation (\ref{rtz}) is automatically 
satisfied without
giving any further independent restriction. This can be shown as follows.
Note the Bianchi identity,
\eq
\nabla_\mu H^\mu\!_\nu =0\,,
\label{bianchi}
\eqx
where $H^\mu\!_\nu$ is the tensor appearing in the usual form of 
the equation of motion
\eq
H^\mu\!_\nu\equiv 
R^\mu\!_\nu - {1\over 2} \delta^\mu\!_\nu R -8\pi G_5 {\cal T}^\mu\!_\nu
=0
\eqx
with ${\cal T}^\mu\!_\nu$ including the contributions from the
 gauge fields and the cosmological constant. The equation 
(\ref{bianchi}) is a mathematical identity once  the solution (\ref{maxsol})
of the Maxwell
equations is used, which insures the covariant conservation of 
the tensor ${\cal T}^\mu\!_\nu$. 
Then $\nu=\tau$ component of the Bianchi identity
 can be written as
\eq
\partial_z  (\sqrt{-g}H^z\!_\tau)
+ 
\partial_\tau  (\sqrt{-g}H^\tau\!_\tau)
-\sqrt{-g}\Gamma^\alpha_{\tau\beta} H^\beta\!_\alpha =0\,.
\eqx
Note that $H^z\!_\tau$ is  $O(z/\tau)\,\,(=O(v/\tau^{2\over3}))$ in its 
leading order. 
Assuming (\ref{gtt})-(\ref{gxx}) are solved in their leading order, 
$H^z\!_z=H^y\!_y=H^x\!_x=O(v/\tau^{2\over3})$ at most. 
The terms involving $H^z\!_z=H^y\!_y=H^x\!_x$ in (\ref{bianchi}) 
have an extra $\tau$ derivative and, therefore, 
are $v/\tau^{2\over3}$ higher order
than the leading terms of  $H^z\!_\tau$.
Hence by setting
those higher order terms to zero, one is left with
\eq
\partial_z  \left(\sqrt{{-g}}H^z\!_\tau\right) =0\,.
\eqx  
Fixing the integration constant to zero by the asymptotic condition,
one does have the relation $H^z\!_\tau =R^z\!_\tau=0$. Thus the
off-diagonal equation 
automatically follows from the remaining part of the equations.

Let us emphasize that the above procedure gives a way to obtain the
form of asymptotic geometry only for large proper-times from the
corresponding static solution. The resulting solution is, however, not
an  exact solution for smaller $\tau$. The sub-asymptotic form of the
metric does not seem to be any longer linked to the static
solution. In a way this is not surprising since the nonlinear
evolution includes effects of viscosity and even more pronounced
deviations from perfect fluid hydrodynamics for small
proper-times. What seems to be more surprising is the fact that any
such correspondence between static and evolving boost-invariant
solutions exists at all.

\section{Thermodynamics}

Let us now analyze the thermodynamic  properties of the
evolving geometry derived in the previous section.

Let us first evaluate the entropy per unit  volume. The boundary 
volume element 
here is given by  
\eq
dV=\tau dy dx^1 dx^2\,.
\eqx
We will use just the extrapolated static formulas since in
any case the asymptotic geometry is not valid {\em at} the horizon but
rather in the scaling limit $\tau \to \infty$ with $v$ fixed.
The location of the extrapolated horizon is thus obtained by finding a smaller
positive root of
\eq
1-a v_h^4 +\f{q^2}{12} v_h^6=0\,,
\eqx
so the entropy density is given by 
\eq
s=\f{S}{V} = \f{N_c^2}{2\pi}\cdot \f{1}{z_h^3} =
\f{N_c^2}{2\pi} \cdot \f{1}{v_h^3\tau} 
\eqx
while the temperature is
\eq
T=\f{|h'(z_h)|}{4\pi} =\f{1-\f{q^2}{8a}v_h^2}
{\pi v_h (1 - \f{q^2}{12 a}v_h^2)} \tau^{-\f{1}{3}}\,.
\eqx

In order to find the energy density $\eps(\tau)$ one has to pass to
Fefferman-Graham coordinates defined by
\eq
ds^2=\f{1}{\zfg^2} \left( -e^{a_{FG}(\zfg,\tau)} d\tau^2+\ldots
    +d\zfg^2 \right)
\eqx
and read off $\eps(\tau)$ from
\eq
\label{epsformula}
\eps(\tau) =\f{N_c^2}{2\pi^2} \lim_{\zfg \to 0} \f{-a_{FG}(\zfg,
  \tau)}{\zfg^4}\,. 
\eqx
We have to redefine the $z$ variable as
\eq
\label{redef}
z=\zfg \left(1-\gamma \f{\zfg^4}{\tau^{\f{4}{3}}}+\ldots \right)
\eqx
and requiring 
\eq
\f{\left(1-a \f{z^4}{\tau^{\f{4}{3}}} +\f{q^2}{12} \f{z^6}{\tau^2}\right)^{-1}
dz^2}{z^2}=\f{d\zfg^2}{\zfg^2}  
\eqx
fixes the coefficient $\gamma=a/8$. At this order and in the scaling limit
one does not need to redefine $\tau$. Applying the redefinition
(\ref{redef}) to the $g_{\tau\tau}$ component of the metric we use
(\ref{epsformula}) to get 
\eq
\eps(\tau)=\f{N_c^2}{2\pi^2}
\f{a-2\gamma}{\tau^{\f{4}{3}}}=\f{N_c^2}{2\pi^2} \cdot \f{3}{4}a \cdot
\tau^{-\f{4}{3}}  =\f{3 N_c^2}{8\pi^2 v_h^4 (
1 - \f{q^2}{12 a}v_h^2
)} \tau^{-\f{4}{3}}\,.
\eqx
By the similar procedure for the other component of the metric,
 one finds that $p=\epsilon/3$.

Finally the charge and chemical potential can be found from the
behavior of the gauge field. 
Let us first work out the charge density. 
The displacement is evaluated as
\eq
D^{z\tau}= {1\over 16\pi G_5} \sqrt{-g} F^{\tau z} ={N_c^2\over 8\pi^2} q\,,
\eqx
which is constant.  The boundary charge density
is then defined by
\eq
\rho = \int dy dx^1 dx^2 D^{z\tau}/V =   {N_c^2\over 8\pi^2} q/\tau
\eqx
showing the expected $\tau$ dependence\footnote{The boundary 
charge density and the chemical potential 
can be determined by studying the boundary behavior of
the gauge fields using AdS/CFT dictionary. The results are the same as
the ones from the bulk method used here.}.
The chemical potential can be given as the difference between
the horizon and the boundary values of the Coulomb potential
\eq
\mu = A_\tau(z_h)-A_\tau (z=0) = \f{1}{2} \cdot   
{qv_h^2\over \tau^{\f{1}{3}}}\,.
\eqx 
One may check that the Gibbs potential $\Omega$ satisfies
the relation,
\eq
\Omega/V = -p= \epsilon - T s - \mu\rho\,.
\eqx
Also the energy density can be written as
\eq
\epsilon(s,
\rho)= \f{3 s^{\f{4}{3}}}{2 (2\pi N_c)^{\f{2}{3}}} (1+\f{4\pi^2 \rho^2}
{3 s^2})\,,
\eqx
from which one may check that the chemical potential 
$\mu$ is indeed conjugated to $\rho$ by the relation 
$\mu={\partial \epsilon/ \partial \rho}$.
Finally there is the requirement of thermodynamic stability
leading to the condition \cite{Cvetic,Sta,Yamada},
\eq
\rho^2 \le \f{3 s^2}{4\pi^2}\,.
\eqx

\section{The dilaton and electric/magnetic equilibration}

It is interesting to consider turning on the dilaton instead of the
gauge field. Such a setup would correspond to considering
configurations (states) in gauge theory with a nonvanishing
expectation value of $\tr F_{\mu\nu} F^{\mu\nu}$ (note that this is now
the 4D {\em gauge theory} field strength and not the 5D supergravity
field considered before in this paper). This means that such a
configuration of the plasma has 
\eq
\cor{\tr \vec{E}^2} \neq \cor{\tr \vec{B}^2}\,.
\eqx

The corresponding 5D action in the Einstein frame is given by
\begin{eqnarray}
I= {1\over 16\pi G_5}
\int \sqrt{-g} \Bigl(R+ 12 - {1\over 2} g^{\alpha\beta}
\partial_\alpha\phi \partial_\beta\phi\Bigr) 
\,.
\end{eqnarray} 
The supergravity equations of  motion becomes
\begin{eqnarray}
&& R_{\alpha\beta}= -4 g_{\alpha\beta}  +
{1\over 2}\partial_\alpha \phi \partial_\beta \phi
\ ,\\
&& \nabla^2 \phi=0
\,.
\end{eqnarray}
As in Ref.~\cite{JP1}, 
we shall use the Fefferman-Graham form of coordinate by setting
$D=0$ this time while allowing a nontrivial 
$C(z,\tau)= c(v)+ O(1/\tau^\natural)$ with the general scaling variable
$v=z/\tau^{s/4}$ with $0<s <4$. 
Also in the boost invariant scaling 
limit, the scalar field behaves
\eq
\phi(z,\tau)=\varphi (v)+O(1/\tau^\natural)\,.
\eqx

The scalar field equation in the leading order can be 
integrated leading to
\eq
\varphi'= \f{k v^3}{ e^{\f{a+b}{2}+c}}\,.
\eqx

Using this expression, 
the remaining  Einstein equations  become
\begin{eqnarray}
&&
-2v (a'+b'+ 2 c')+ v^2 ( (a')^2 + (b')^2 + 2(c')^2 +
2 (a'' + b'' + 2 c''))\nonumber\\
&& = -2 \f{k^2 v^8}{e^{a+b+2c}}
\ ,
\label{eq1}
\\
&&
-2v ( 4 a'+b'+ 2 c')+ v^2 ( (a')^2 + a'(b' + 2c') +
2 a'' 
)= 0
\ ,\label{eq2}\\
&&
-2v (  a'+ 4 b'+ 2 c')+ v^2 ( (b')^2 + b'(a' + 2c') +
2 b'' 
)=  0
\ ,\label{eq3}\\
&&
-2v ( a'+b'+ 5 c')+ v^2 ( 2(c')^2 + c'(a' + b') +
2 c'' 
)=  0
\ ,\label{eq4}\\
&&
2v \bigl( \f{4 (b'-a')}{s} -b'-2 c'\bigr)- v^2 ( (b')^2 - a'(b' + 2c') +
2 (c')^2+ 2b'' +4 c''
)\nonumber\\
&&= 2 \f{k^2 v^8}{e^{a+b+2c}}
\ ,\label{eq5}
\end{eqnarray}
Interestingly these equations can be solved exactly. The steps 
are as follows. First we add 
 (\ref{eq1}) and (\ref{eq5}) to cancel out the scalar contributions.
We then linearly combine the resulting expression with 
(\ref{eq2}) such that all the second derivatives cancel out. 
The final result becomes
\begin{eqnarray}
 (4-3s)a' +(s-4)b'  + 2s c'=0\,.
\end{eqnarray}
Using the boundary condition $a(0)=b(0)=c(0)=0$, this is solved by
\begin{eqnarray}
a=M-2m\,, \ \ b= M+ (2s-2)m\,,\ \  c= M+ (2-s)m\,.
\end{eqnarray}
with $M(0)=m(0)=0$.
Inserting these expressions
into Eqs. (\ref{eq2})-(\ref{eq4}), 
one finds that the three equations  are reduced to 
\begin{eqnarray}
&&
-7v M' + v^2 (  2(M')^2 +  M'') = 0
\,,
\label{eq11}\\
&&
-3v m' + v^2 (  2 m'M' +   m'') = 0
\ .\label{eq12}
\end{eqnarray}
These are solved by \cite{JP1} 
\begin{eqnarray}
&&
M=\f{1}{2} \ln(1-\Delta^2 v^8)
\,,
\label{eq21}\\
&&
m= \f{1}{4\Delta} \ln\left({1-\Delta v^4\over 1+\Delta v^4 }\right)
\,.\label{eq22}
\end{eqnarray}
What remains is to satisfy (\ref{eq1}) with the above expression of
$M$ and $m$, which leads to
\eq
\Delta^2 =\f{3s^2 -8s +8}{24} + \f{k^2}{96}\,. 
\eqx 

Then finally computing $R_{\mu\nu\alpha\beta}R^{\mu\nu\alpha\beta}$ \cite{JP1},
one can check that it is nonsingular only for  $k=0$ and $s=4/3$. Therefore
one concludes that the evolving geometry 
with a nontrivial scalar field 
 is not allowed for the late time.

The above result shows that fluctuations of electric and magnetic
modes tend to equilibrate very fast. Perhaps faster than power-like
behavior but note that above result only  shows  that 
the scalar field should approach zero faster than
$1/\tau^{\natural}$ with $\natural > 0$ for the finite scaling
coordinate $v$. 

This is analogous
to the stability property of the evolving 
perfect fluid geometry with respect to small scalar perturbations
\cite{JP2} in the linearized theory. These are in effect quasinormal
modes which exhibit exponential decay. The above analysis extends this
result to states far away from electric/magnetic equilibration.

\section{Conclusions}

In this note, we obtain the asymptotic evolving 
boost-invariant geometry involving conserved R-charge. Thereby the
boost-invariant  
late time dynamics of strongly coupled ${\cal N}=4$ Super Yang-Mills
with R-charge turned on are studied via
the AdS/CFT correspondence. The result shows that
the boost invariant late time state has necessarily to be
in the perfect-fluid hydrodynamic regime even including the R-charge.


The asymptotic large proper-time boost-invariant evolving geometry can
be seen to arise from the corresponding static solution by
substituting the scaling variable $v$ for $z$ in the coefficient
functions. This property arises due to the specific form of the Ricci tensor in
the scaling limit which closely mirrors the Ricci tensor for the
analogous static solution. Let us note, however, that the asymptotic
geometry is not an {\em exact} solution of the Einstein equations and
sub-asymptotic corrections exist. These corrections are important as
they encode specific time-dependent dynamical effects. 

We also discuss the 5D  Einstein-scalar theory to show 
that the late time boost invariant geometry with a nontrivial scalar
field is not allowed.  
In the gauge theory side  this implies that a difference between 
electric and magnetic modes does not survive into the asymptotic
proper-time regime. 


\bigskip 
\vskip 1cm

\noindent{}{\bf Acknowledgments.} 
We are grateful to Andreas Karch, Pavel Kovtun, Dam Son and 
Larry Yaffe for useful discussions and conversations.
The work of DB is supported in part by
KOSEF ABRL R14-2003-012-01002-0 and KOSEF SRC CQUeST R11-2005-021.
This work was initiated during
the program `From RHIC to LHC: Achievements and Opportunities' at the
Institute of Nuclear Theory, Seattle. RJ would like to thank the INT
for hospitality.
RJ was supported in part by Polish Ministry of Science and Information
Society Technologies grants 1P03B02427 (2004-2007), 1P03B04029
(2005-2008) and RTN network ENRAGE MRTN-CT-2004-005616.

\appendix

\subsection*{Appendix. Expressions for the scaling limit of $R^\al\!_\bt$}

\def\theequation{a
.\arabic{equation}}

Here we quote expressions for the diagonal components of the Ricci
tensor used in the main text:
\eqn
R^\tau\!_\tau &=& -\f{1}{4} e^{-d(v)} (16+v^2(2a''(v)+a'(v)^2)+2v
(b'(v)+d'(v)-4a'(v)) + \nonumber \\
&& +v^2 (b'(v)+d'(v))a'(v) )\\
R^z\!_z &=& -\f{1}{4}e^{-d(v)} ( 16+v^2(2a''(v)+a'(v)^2)+
v^2(2b''(v)+b'(v)^2) + \nonumber \\
&& -2v (a'(v)+b'(v)-4d'(v)) -v^2 (b'(v)+a'(v))d'(v)) \\
R^y\!_y &=& -\f{1}{4}e^{-d(v)} ( 16+v^2(2b''(v)+b'(v)^2) -2v
(a'(v)-d'(v)+4b'(v)) +\nonumber \\
&& -v^2 (d'(v)-a'(v))b'(v)) \\
R^x\!_x &=& -\f{1}{2}(8+vd'(v)-va'(v)-vb'(v))\,.
\eqnx

\end{document}